\documentclass[a4paper,12pt]{article}
\usepackage[utf8]{inputenc}

\usepackage{amsfonts}
\usepackage{amssymb}
\usepackage{graphicx}
\usepackage{amsmath}
\usepackage{enumerate}
\usepackage{color}
 \usepackage{multirow}
\usepackage{float}

\RequirePackage{mathrsfs} \RequirePackage[sc]{mathpazo}
\RequirePackage{wasysym} \RequirePackage{setspace}\textheight=650pt
\title{ Statistical Physics  of  $3D$  Hairy Black Holes }
\author{\hspace{-1cm}A. Belhaj$^{1,2}$, M. Chabab$^2$, H. EL Moumni$^2$, K. Masmar$^2$, M.  B. Sedra$^{3,4}$ \\
\\
{\small \hspace{-1cm}$^{1}$D\'epartement de Physique, Facult\'e
Polydisciplinaire, Universit\'e Sultan
 Moulay Slimane, B\'eni Mellal,  Morocco. } \\
{\small \hspace{-1cm}$^{2}$High Energy Physics and Astrophysics Laboratory, FSSM,
 \small Cadi Ayyad University, Marrakesh, Morocco.
} \\
{\small \hspace{-1cm}$^{3}$ D\'{e}partement de Physique, LHESIR, Facult\'{e} des
Sciences, Universit\'{e} Ibn Tofail,
 K\'{e}nitra, Morocco.} \\
 {\small \hspace{-2 cm}$^{4}$Universit\'e Mohammed Premier,
  Ecole Nationale des Sciences Appliqu\'ees, BP : 3, Ajdir, 32003, Al Hoceima, Morocco.}
 }
\date{ }

\begin{document}


\maketitle


\abstract{ We investigate  the  statistical behaviors  of   3D hairy
black holes in the presence of  a  scalar field.  The present study
is made in terms  of two relevant parameters:   rotation parameter
$a$ and   $B$ parameter related to the scalar field. More precisely,
we compute  various statistical quantities including the partition
function for non-charged  and charged  black hole solutions.   Using
a  partition function calculation, we show that the probability is
independent of $a$ and $B$ parameters}.

\newpage
Recently, many  effects have been devoted  to study  thermodynamic
behaviors of black holes in lower and higher dimensions
 \cite{30,a1,a2,4,5,50,our,our1,our3d,our3,our4,ou5,our6,x7,x9,14}.  For
 certain
systems, the equation of states have been  worked out  sharing
similarities with Van der Waals P-V  systems. In four dimensions for
instance, RN-AdS black holes with spherical  geometries  have been
extensively investigated \cite{KM,GKM}.  More precisely, it has been
remarked that there is  a nice interplay between the behaviors of
the RN-AdS black hole systems which  has been explored in many
works. The P-V criticality, the Gibbs free energy, the first order
phase transition and the behavior near the critical points are
associated  with  the statistical liquid-gas systems. In particular,
the critical behaviors of charged RN-AdS black holes in arbitrary
dimensions of the spacetime have been investigated  \cite{our}.

On the other hand, a particular interest has been put on the three
dimensional case corresponding to  the  BTZ black hole whose
critical behaviors are associated with the ideal gas ones
\cite{our,our3d}. More recently, a novel exact rotating black hole
solution in (2+1)-dimensional gravity with a non-minimally coupled
scalar field has been studied using   an appropriate metric ansatz
\cite{ZXZ, XZ}. In  this way, critical behaviors of a class of such
black holes
 has  been investigated.  Interpreting the  cosmological constant
as a thermodynamic pressure and its conjugate quantity as a volume,
 the corresponding equation of state has been established. In a generic region of the
 corresponding moduli space, these  black  holes behave like a Van der Waals system \cite{our3d}.

The aim of this  work   is  to  contribute  to these activities by
studying the statistical behaviors of   3D  hairy black holes. In
particular, we compute  various  statistical quantities including
the partition function for non-charged  and charged solutions.  This
study is made  in terms of two parameters $B$ and $a$. These
parameters are associated with  the scalar field and the angular
momentum respectively.  Using  a  partition function  calculation,
we reveal  that the probability is independent of  such parameters.

To start we   reconsider the study of  the statistical physics
 of  $3D$-dimensional gravity with a
non-minimally coupled scalar field.   This   black hole  solution is
known as hairy black hole in three dimensions. In the absence of the
Maxwell   gauge fields, this model  can be  described  by the
following action $\cite{ZXZ}$
\begin{equation}
\mathcal{I_R}=\frac{1}{2}\int d^3x \sqrt{-g}\left[R -g^{\mu\nu}\nabla_\mu\phi\nabla_\nu\phi-\xi R\phi^2-2V(\phi)\right]
\end{equation}
where $\phi$ is  the  dynamical scalar field.  For simplicity
reason, we  consider a particular   situation where  the coupling
and the gravitational constants  are fixed to   $\xi=\frac{1}{8}$
and $\kappa=8\pi G=1$ respectively.   In this way, the  solution
takes the following form
\begin{equation}\label{metric}
ds^2=-f(r) dt^2+\frac{1}{f(r)}
dr^2+r^2\left(d\theta+\omega(r)dt\right)^2.
\end{equation}
It is noted that  the  functions $f$ and $w$, controlling such a
solution, read as
\begin{equation}\label{fr}
f(r)=3\beta+\frac{2B\beta}{r}+\frac{(3r+2B)^2 a^2}{r^4}+\frac{r^2}{\ell^2}\end{equation}
\begin{equation}\label{omega}
\omega(r)=-\frac{(3r+2B)a}{r^3}
\end{equation}
where  $\beta$ is identified with $-\frac{M}{3}$. In these
equations, $a$ describes the rotating  angular momentum parameter.
While, the parameter $B$ is linked  to the dynamical  scalar field.

For later use, an explicit form of the potential will be needed. In
the present study, we  explore   the    potential $V(\phi)$ proposed
in \cite{XZ} given by the following equation
\begin{equation}
V(\phi)=\frac{1}{512} \left(\frac{a^2 \left(\phi ^6-40 \phi ^4+640 \phi ^2-4608\right)
   \phi ^{10}}{B^4 \left(\phi ^2-8\right)^5}+\phi ^6 \left(\frac{\beta
   }{B^2}+\frac{1}{\ell^2}\right)+\frac{1024}{\ell^2}\right).
\end{equation}
Having fixed the metric backgrounds,  we  move now to study
thermodynamical behaviors of such black hole solutions. More
precisely, we compute the entropy considered as the most important
thermodynamical quantity. In $3D$, the calculation produces  the
following  entropy function
\begin{equation}\label{6}
s=4\pi r_+
\end{equation}
where $r_+$ is the horizon radius obtained by solving   $f(r)=0$.
Taking this entropy function, we can derive  many statistical
quantities including the  temperature, the  momentum and  the
specific  heat capacity.   In order to obtain such quantities, we
should first determine the black hole mass.   Indeed,  the equations
(\ref{6}) and (\ref{metric}) give the  entropy-dependent mass
  \begin{equation}
  M=\frac{48 \pi ^2 a^2 (8 \pi  B+3 s)}{s^3}+\frac{3 s^3}{16 \pi
   ^2 \ell ^2 (8 \pi  B+3 s)}.
  \end{equation}
The corresponding   numerical behaviors are presented in figure 1.
  \begin{center}
\begin{figure}[!ht]
\center
\begin{tabbing}
\hspace{10cm}\=\kill
\includegraphics[scale=.9]{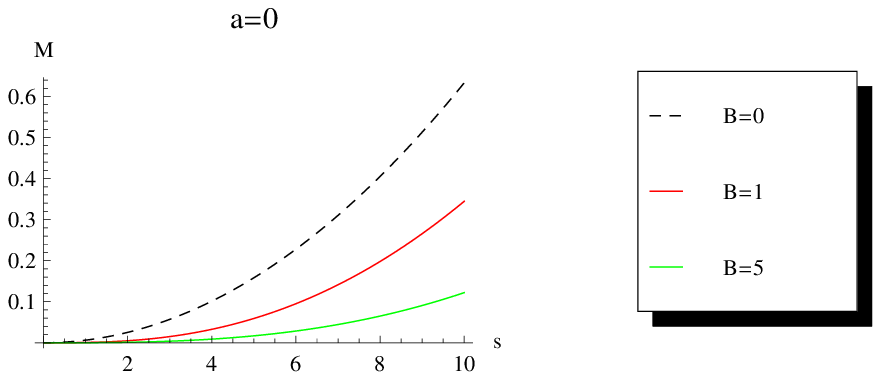} \> \includegraphics[scale=.9]{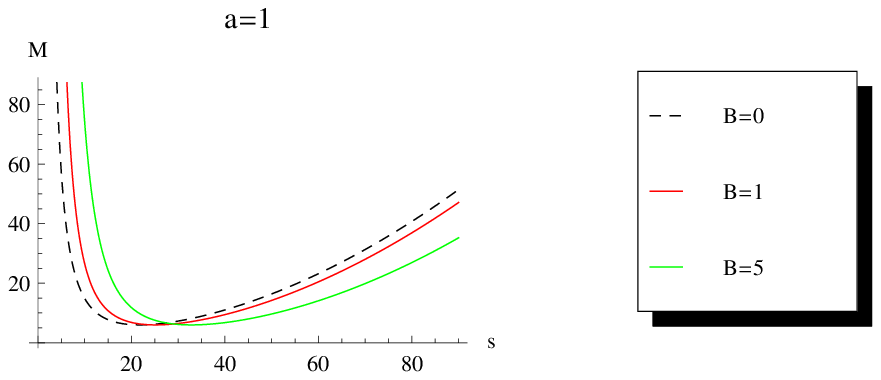}
\end{tabbing}
\caption{Plots of black hole mass for  $\ell = 1$. }
\label{fig1:fig2}
\end{figure}
\end{center}

It follows from this figure  that  the   scalar charge  $B$
decreases the mass of the 3D black hole.  More precisely, the
rotation parameter modifies the thermodynamical behavior of the mass
and  the temperature in terms of  the  entropy. In such a solution,
the mass function  shows a minimum. The corresponding moduli space
contains two relevant regions associated with the asymptotic
behaviors of the entropy. Indeed, in the first region corresponding
to small values, the mass of the  3D black holes increases with  the
$B$ parameter. However, in the large limit values associated with
second region, the mass decreases with  the $B$ parameter. Similar
behaviors occur in the corresponding   black hole temperature. It is
recalled that the later can be obtained by using the following
relation
 \begin{equation}\label{7}
 T=\left(\frac{\partial M}{\partial s}\right).
\end{equation}
Exploring eq.(\ref{7}), the computation  can   give  the black hole
temperature in terms of the entropy  function.  The obtained
relation reads as
  \begin{equation}\label{temperature}
  T=\frac{9 s^2 (4 \pi  B+s)}{8 \pi ^2 \ell ^2 (8 \pi  B+3
   s)^2}-\frac{288 \pi ^2 a^2 (4 \pi  B+s)}{s^4}.
  \end{equation}
This temperature  function is illustrated in figure 2.
   \begin{center}
\begin{figure}[!h]
\center
\begin{tabbing}
\hspace{10cm}\=\kill
\includegraphics[scale=.9]{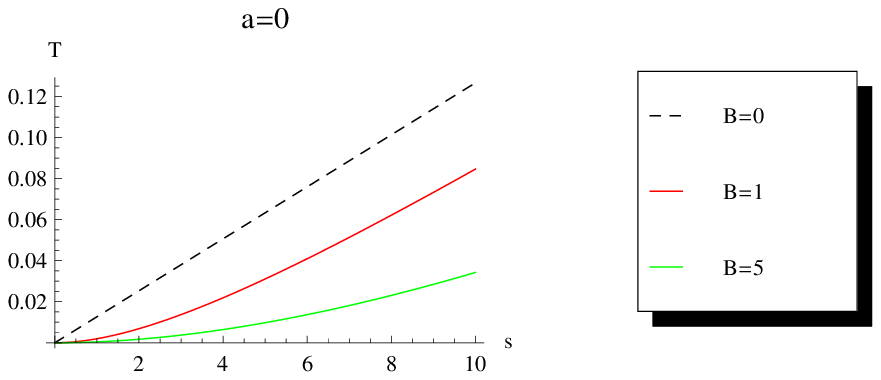} \> \includegraphics[scale=.9]{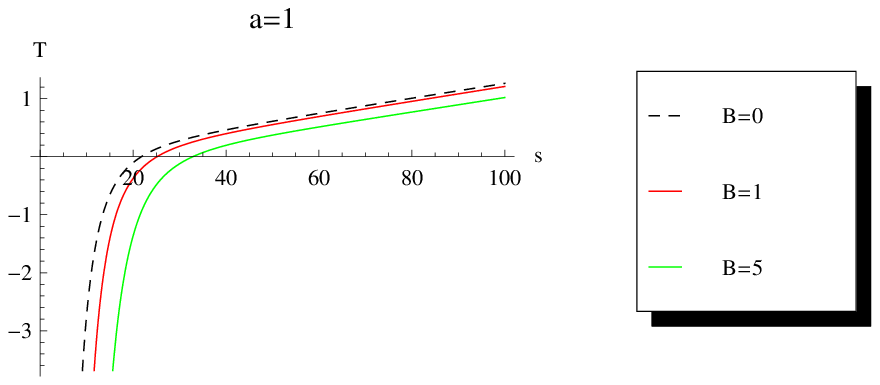}
\end{tabbing}
\caption{Plots of black hole temperature for $\ell = 1$. }
\label{fig1:fig2}
\end{figure}
\end{center}

From this figure, it has been observed similar behaviors appearing
in  the mass  term of the black hole. However, the novelty is
associated with the negative value of the temperature.

To complete the statistically investigation of the such  3D black
holes,  we compute the specific heat. This quantity can  be
obtained from the following equation
   \begin{equation}\label{9}
  C =T\left(\frac{\partial s}{\partial T}\right).
\end{equation}
Using Eq.$(\ref{temperature})$, we  get  the specific heat function
  \begin{equation}
  C=\frac{s (4 \pi  B+s) (8 \pi  B+3 s) \left(s^6-256 \pi ^4 a^2 \ell ^2 (8 \pi  B+3
   s)^2\right)}{256 \pi ^4 a^2 \ell ^2 (8 \pi  B+3 s)^3 (16 \pi  B+3 s)+s^6 \left(64
   \pi ^2 B^2+24 \pi  B s+3 s^2\right)}.
  \end{equation}
This function is presented in figure 3.
    \begin{center}
\begin{figure}[!h]
\center
\begin{tabbing}
\hspace{10cm}\=\kill
\includegraphics[scale=.9]{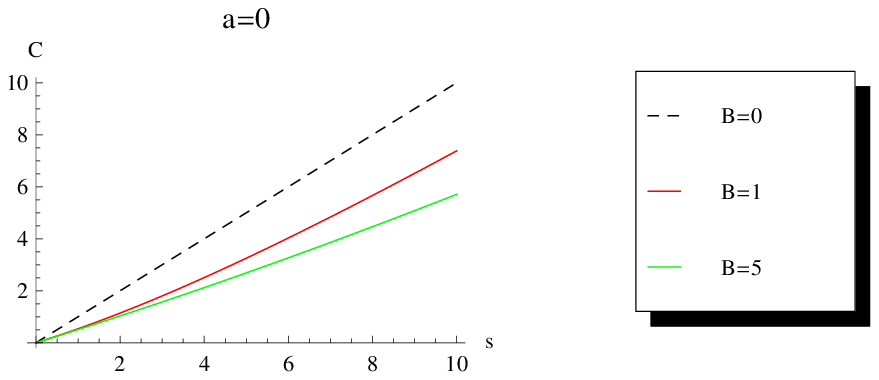} \> \includegraphics[scale=.9]{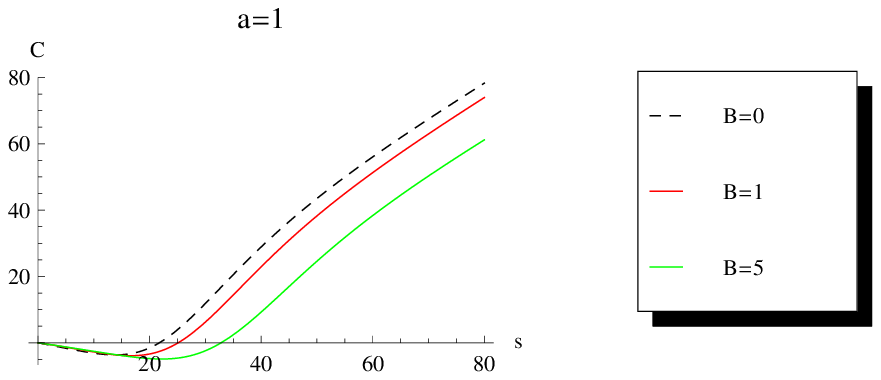}
\end{tabbing}
\caption{Plots of black hole heat capacity for $\ell = 1$. }
\label{fig1:fig2}
\end{figure}
\end{center}

It is recalled that  the sign of  the specific heat can  determine
the thermodynamical  stability of the  black hole. The positivity of
such a quantity   ensures the  stable equilibrium \cite{14}. It
follows from the  figure 3 that the rotating parameter leads to an
instability  solution. Indeed, it has been observed that similar
 behaviors  appear  in non rotating solution. This means that  the heat capacity decreases with
 scalar charge $B$. In fact,  it is
  still always positive explaining  that the black hole solution is stable.
However, the  introduction of the rotation parameter modifies  the
black hole behaviors and perturbs
 its   the stability.   Moreover, the heat capacity shows a minima
and takes  negative values in the first  region  associated with
small values of the entropy.

It is  observed that  in the vanishing limit of $a$ with
$\beta=-\frac{B^2}{\ell^2}$, the  specific heat reduces  to
\begin{equation}
C=8\pi\ell\sqrt{\frac{M}{3}}=s
\end{equation}
recovering  the result  reported in \cite{Ir}.

The obtained  quantities will be useful for discussing the most
important function used in statistical physics.  In particular, we
compute  the  partition function.  The calculation gives the
following form
\begin{eqnarray}
Z&=&\exp\left(-\frac{F}{ T}\right)\\
&=&\exp\left[\frac{2 \pi  r^4 \left(\frac{\frac{16 B^3 (B+3
   r)}{(2 B+3 r)^2}+9 r^2}{\ell ^2}-\frac{27 a^2
   (8 B+9 r)}{r^3}\right)}{81 (B+r)
   \left(\frac{r^6}{\ell ^2 (2 B+3
   r)^2}-a^2\right)}\right]
\end{eqnarray}
where $F$ is the free energy  expressed as follows
\begin{eqnarray}
F&=&-\int s dT\\
&=& \frac{24 a^2 B}{r^3}+\frac{27 a^2}{r^2}-\frac{16
   B^4}{9 \ell ^2 (2 B+3 r)^2}-\frac{16 B^3 r}{3
   \ell ^2 (2 B+3 r)^2}-\frac{r^2}{\ell ^2}.
\end{eqnarray}
The probability can be derived  from the partition function. The
computation  leads to
\begin{eqnarray}\label{ppp}
p=e^{-e^{-4 \pi  r_+}}=e^{-e^{s}}.
\end{eqnarray}
It is observed  that this  quantity  does not depend neither on  the
 scalar charge $B$  nor the rotation parameter
 $a$.

Having discussed the non charged 3D black hole solution, now we
investigate the case of
 the charged back holes.  Introducing the  Maxwell gauge field, the
 corresponding action, as given in  \cite{ZXZ}, takes the following
 form
\begin{equation}
\mathcal{I_{RQ}}=\frac{1}{2}\int d^3x \sqrt{-g}\left[R -g^{\mu\nu}
\nabla_\mu\phi\nabla_\nu\phi-\xi
R\phi^2-2V(\phi)-\frac{1}{4}F_{\mu\nu}^{\mu\nu}\right].
\end{equation}
For simplicity reason, the coupling constant is fixed to
$\frac{1}{8}$ and the self coupling  potential is given by
\begin{equation}
V(\phi)=\frac{2}{\ell^2}+\frac{1}{512}\left[\frac{1}{\ell^2}+
\frac{\beta}{B^2}+\frac{Q^2}{9 B^2}\left(1-\frac{3}{2}ln\left(\frac{8 B}{\phi^2}\right)
\right)\right]\phi^6+\mathcal{O}(Q^2 a^2 \phi^8)
\end{equation}
where $Q$ is the infinitesimal electric charge.  The parameter
$\beta$  takes the following form \begin{equation}
\beta=\frac{1}{3}\left(\frac{Q^2}{4}-M\right).
\end{equation}
For this solution, the metric background $(\ref{fr})$  get modified
and becomes as follows
\begin{equation}
f(r)=3\beta-\frac{Q^2}{4}+\left(2\beta-\frac{Q^2}{9}\right)\frac{B}{r}-Q^2\left(\frac{1}{2}+
\frac{B}{3r}\right)ln(r)+\frac{(3r+2B)^2a^2}{r^4}
+\frac{r^2}{\ell^2}+\mathcal{O}(a^2Q^2)\end{equation} where $\omega$
is still given by $(\ref{omega})$. Using a  similar analysis, as
used in  the previous section, we  can  obtain  the mass in terms of
the entropy. The calculation produces  the following mass equation
\begin{equation}\label{MM}
M_Q=\frac{32 \pi ^3 \ell ^2 \left(72 \pi  a^2 (8 \pi  B+3 s)^2+B Q^2 s^3\right)+24
   \pi ^2 Q^2 s^3 \ell ^2 (8 \pi  B+3 s) (\log (4 \pi )-\log (s))+9 s^6}{48 \pi
   ^2 s^3 \ell ^2 (8 \pi  B+3 s)}.
\end{equation}
This thermodynamical quantity is plotted in figure 4.
\begin{center}
\begin{figure}[!h]
\center
\begin{tabbing}
\hspace{10cm}\=\kill
\includegraphics[scale=.9]{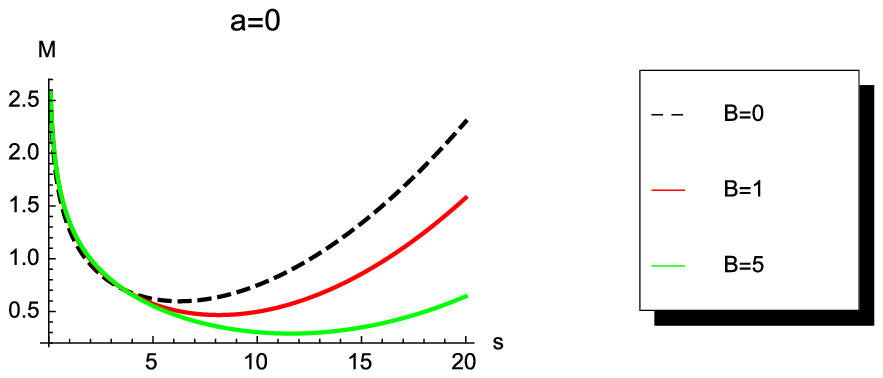} \> \includegraphics[scale=.9]{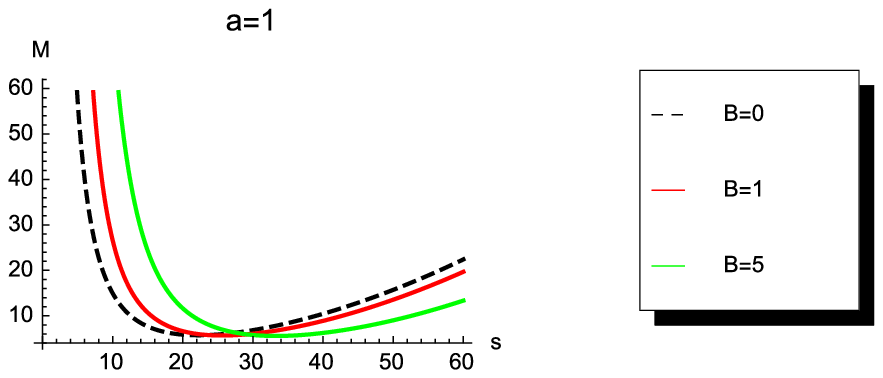}
\end{tabbing}
\caption{Plots of black hole mass for $\ell=Q = 1$. }
\label{fig1:fig2}
\end{figure}
\end{center}
It is observed that the introduction of the rotating parameter leads
to certain   convergences  in the curves. This can bee seen from the
presence of  the $a^2$  term in the mass equation.  Using Eq.
$(\ref{7})$ and $(\ref{MM})$, the temperature, for the charged
solution, can take the following form
\begin{equation}\label{TT}
T_Q=\frac{32 \pi ^3 \ell ^2 \left(72 \pi  a^2 (8 \pi  B+3 s)^2+B Q^2 s^3\right)+24
   \pi ^2 Q^2 s^3 \ell ^2 (8 \pi  B+3 s) (\log (4 \pi )-\log (s))+9 s^6}{48 \pi
   ^2 s^3 \ell ^2 (8 \pi  B+3 s)}.
\end{equation}

\begin{center}
\begin{figure}[!h]
\center
\begin{tabbing}
\hspace{10cm}\=\kill
\includegraphics[scale=.9]{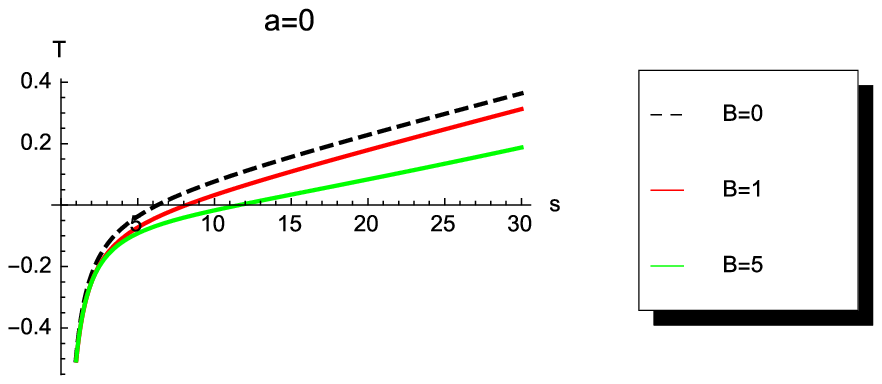} \> \includegraphics[scale=.9]{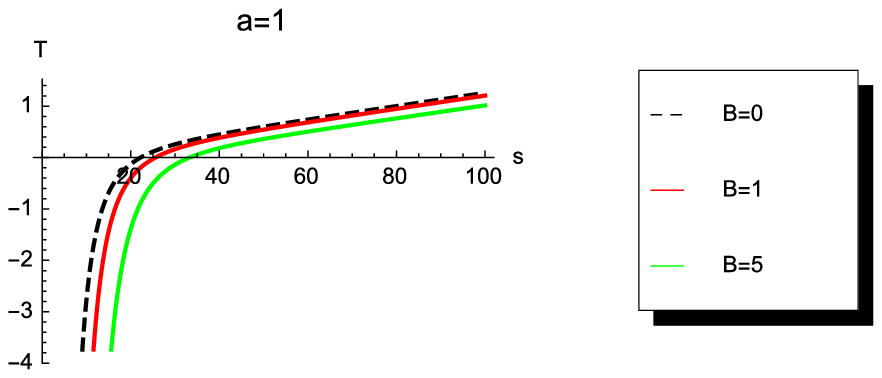}
\end{tabbing}
\caption{Plots of black hole temperature for $\ell=Q = 1$. }
\label{fig1:fig2}
\end{figure}
\end{center}
 Eq. $(\ref{9})$ and $(\ref{TT})$  can be used  to compute  the heat capacity. Indeed, it  becomes

\begin{equation}\textstyle
C_Q=\frac{s (4 \pi  B+s) (8 \pi  B+3 s) \left(9 s^6-4 \pi ^2 \ell ^2 \left(576 \pi
   ^2 a^2 (8 \pi  B+3 s)^2+Q^2 s^3 (16 \pi  B+9 s)\right)\right)}{4 \pi ^2 \ell
   ^2 \left(576 \pi ^2 a^2 (8 \pi  B+3 s)^3 (16 \pi  B+3 s)+Q^2 s^3 \left(512
   \pi ^3 B^3+576 \pi ^2 B^2 s+240 \pi  B s^2+27 s^3\right)\right)+9 s^6
   \left(64 \pi ^2 B^2+24 \pi  B s+3 s^2\right)}.
\end{equation}

\begin{center}
\begin{figure}[!h]
\center
\begin{tabbing}
\hspace{10cm}\=\kill
\includegraphics[scale=.9]{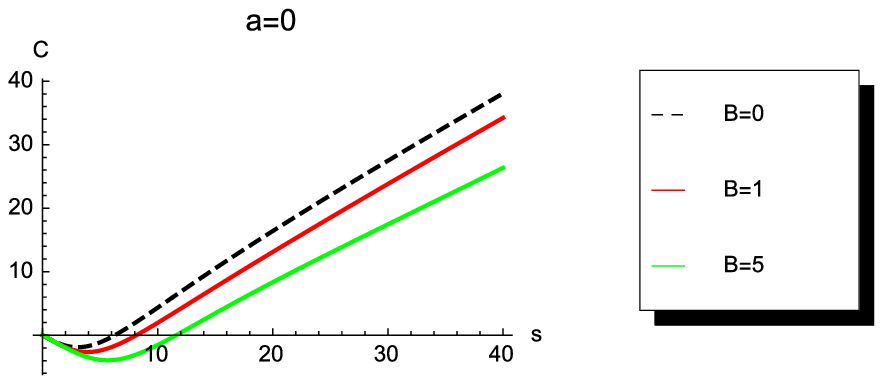} \> \includegraphics[scale=.9]{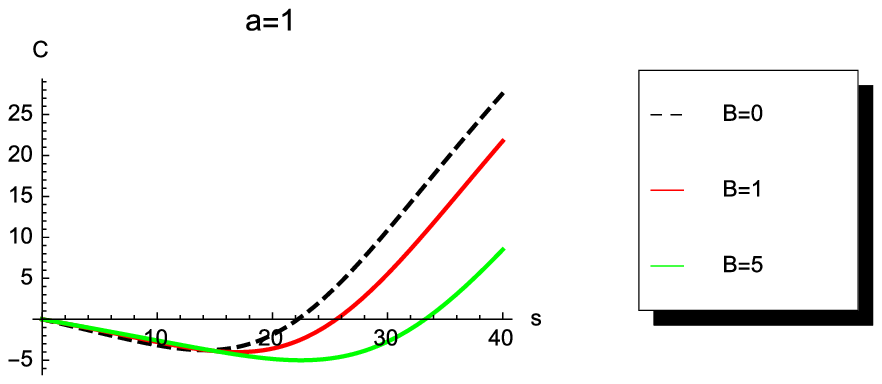}
\end{tabbing}
\caption{Plots of black hole heat capacity for  $\ell =Q= 1$. }
\label{fig1:fig2}
\end{figure}
\end{center}

From   all these   figures, it is observed that   the charged black
hole presents
 an instability. However,   the rotation parameter and the  scalar
 charge can be used to reduce  such an  instability.

To compute the  corresponding  statistical quantities, we keep  the
same analysis used for the uncharged case. Indeed, the  free energy
function reads
\begin{equation}
F_Q= -\frac{1}{2 \ell ^2}\left(-\frac{48 a^2 B \ell ^2}{r^3}-\frac{54 a^2 \ell ^2}{r^2}+\frac{2 B
   (B+3 r) \left(16 B^2-3 Q^2 \ell ^2\right)}{9 (2 B+3 r)^2}+Q^2 \ell ^2
   \log (r)+2 r^2\right).
\end{equation}
In this case, the probability is also independent  of  the charge of
the black hole and  is given by  the equation $(\ref{ppp})$.

In this work, we have investigated the   statistical behaviors  of
3D hairy black holes in the presence of  a  scalar field.  The
 study  has been  made in terms  of two relevant parameters:
the rotation parameter $a$ and $B$ parameter related to the scalar
field. In particular, we  have computed  various statistical
quantities including the partition function for non-charged  and
charged  black hole solutions.   It has been shown, using a
partition function computation,  that the probability  is
independent of the relevant  parameters.

\end{document}